\documentclass[useAMS,usenatbib,usegraphicx]{mn2e}
\bibpunct[, ]{}{}{;}{a}{,}{,}
\newcommand{\cs}{c_{\rm s}}
\newcommand{\Msun}{M_{\odot}}
\newcommand{\Menv}{M_{\rm env}}
\newcommand{\Rflat}{R_{\rm flat}}
\newcommand{\Rmid}{R_{\rm mid}}
\newcommand{\Redge}{R_{\rm edge}}
\newcommand{\rc}{r_{\rm c}}
\newcommand{\rout}{r_{\rm out}}
\newcommand{\rhoc}{\rho_{\rm c}}

\newcommand{\Lbol}{L_{\rm bol}}
\newcommand{\Fco}{F_{\rm co}}

\begin{document}

\title[Finite Mass Reservoir Collapse]{The effect of a finite mass 
reservoir on the collapse of spherical isothermal clouds and the 
evolution of protostellar accretion}
\author[E. I. Vorobyov and S. Basu]{E. I. Vorobyov$^{1,2}$\thanks{E-mail:
vorobyov@astro.uwo.ca (EIV); basu@astro.uwo.ca (SB)} and Shantanu
Basu$^{1}$ \\
$^{1}$Department of Physics and Astronomy, University of Western Ontario, 
London, Ontario, N6A 3K7, Canada \\
$^{2}$Institute of Physics, Stachki 194, Rostov-on-Don, Russia}

\date{Submitted November 1, 2004; Accepted April 1, 2005}

\maketitle

\label{firstpage}

\begin{abstract}
Motivated by recent observations which detect an outer boundary for starless 
cores, and evidence for time-dependent mass accretion in the Class 0 and Class I 
protostellar phases, we reexamine the case of spherical isothermal collapse in 
the case of a finite mass reservoir. The presence of a core boundary results in 
the generation of an inward propagating rarefaction wave. This steepens the gas 
density profile from $r^{-2}$ to $r^{-3}$ or steeper. After a protostar forms, the 
mass accretion rate $\dot{M}$ evolves through three distinct phases: (1) an early 
phase of decline in $\dot{M}$, which is a non-self-similar effect due to spatially 
nonuniform infall in the prestellar phase; (2) for large cores, an intermediate 
phase of near-constant $\dot{M}$ from the infall of the outer part of the self-
similar density profile; (3) a late phase of rapid decline in $\dot{M}$ when 
accretion occurs from the region affected by the inward propagating rarefaction 
wave. Our model clouds of small to intermediate size make a direct transition 
from phase (1) to phase (3) above. Both the first and second phase are 
characterized by a temporally increasing bolometric luminosity $\Lbol$, while 
$\Lbol$ is decreasing in the third (final) phase. We identify the period of 
temporally increasing $\Lbol$ with the Class 0 phase, and the later period of 
terminal accretion and decreasing $\Lbol$ with the Class I phase. The peak in 
$\Lbol$ corresponds to the evolutionary time when $50\% \pm 10\%$ of the cloud mass 
has been accreted by the protostar. This is in agreement with the classification 
scheme proposed by Andr\'{e} et al. (\cite{Andre93}). 
We show how our results can be used to explain tracks of envelope mass 
$\Menv$ versus $\Lbol$ for protostars in Taurus and 
Ophiuchus. We also develop an analytic formalism which reproduces the 
protostellar accretion rate. 

\end{abstract}


\section{Introduction}
\label{Intro}

Recent submillimeter and mid-infrared observations suggest
that prestellar cores within a larger molecular cloud
are characterized by a non-uniform radial gas density distribution
(Ward-Thompson et al. \cite{WT}; Bacmann et al. \cite{Bacmann}).
Specifically, a flat density profile in the central
region of size $\Rflat$ is enclosed within a region of
approximately $r^{-1}$ column density profile (and by implication
an $r^{-2}$ density profile) of extent $\Rmid$. Beyond this, a
region of steeper density ($\rho \propto r^{-3}$ or greater) is 
sometimes detected. Finally, at a distance 
$\Redge$, the column density $N$ seems to merge into a background,
and fluctuate about a mean value that is typical for the 
ambient molecular cloud.
The first two regions, of extent $\Rflat$ and $\Rmid$, respectively, are
consistent with models of unbounded isothermal equilibria or
isothermal self-similar gravitational collapse
(e.g. Chandrasekhar \cite{Chandra}; Larson \cite{Larson}; Penston
\cite{Penston}). In either case, the effect of
an outer boundary is considered to be infinitely far away (i.e.
$\Rmid \rightarrow \infty$ in our terminology).
In numerical simulations of
gravitational collapse in which there is a qualitative change in the
physics beyond some radius (e.g. a transition from magnetically
supercritical to subcritical mass-to-flux ratio: Ciolek \&
Mouschovias \cite{Ciolek93}; Basu \& Mouschovias \cite{Basu94}),
the development of
a very steep outer density profile is also seen. Finally, larger scale
simulations of core formation in clouds with uniform background
column density (Basu \& Ciolek \cite{Basu04}) show an eventual
merger into a near-uniform background column density, demonstrating
the existence of $\Redge$. The implication of an outer density
profile steeper than $r^{-3}$ is that there is a finite reservoir
of mass to build the star(s), assuming that the gas beyond $\Redge$ is
governed by the dynamics and gravity of the parent cloud, and thus
does not accrete on to the star(s) formed within the core.

An important constraint of the observations are the actual 
sizes of the cores. For example, in the clustered star formation
regions such as $\rho$ Ophiuchi protocluster, $\Redge \la 5000$ AU, and
$\Redge/\Rflat \la 5$, while in the more extended cores in
Taurus, $5000~{\rm AU} \la \Redge \la 20000$ AU, and $5\la \Redge/\Rflat \la 10$
(see Andr\'e et al. \cite{Andre}; Andr\'e, Ward-Thompson, \& Barsony \cite{Andre2}).
Clearly, only the latter case may approach self-similar conditions.

Once a central hydrostatic stellar core has formed, the mass 
accretion rate is expected to be constant
in isothermal similarity solutions
(Shu \cite{Shu}; Hunter \cite{Hunter2}; Whitworth \& Summers \cite{Whit}).
For example, for the collapse from rest of a singular isothermal 
sphere (SIS) with density profile
$\rho_{\rm SIS} = \cs^2/(2 \pi G r^2)$, where $\cs$ is the isothermal
sound speed, Shu (\cite{Shu}) has
shown that the mass accretion rate ($\dot{M}$) is
constant and equal to $0.975 \, c^3_{\rm s}/G$.
However, two effects can work against a constant $\dot{M}$ 
in more realistic scenarios of isothermal collapse: 
(1) inward speeds in the prestellar
phase are not spatially uniform as in the similarity solutions, and tend
to increase inward, meaning that inner mass shells fall
in with a greater accretion rate; (2) the effect of a finite mass
reservoir will ultimately reduce accretion.
The first effect has been clearly documented in a series of
papers (e.g. Hunter \cite{Hunter2}; Foster \& Chevalier \cite{FC};
Tomisaka \cite{Tomisaka}; Basu \cite{Basu97}; Ciolek \& K\"{o}nigl
\cite{Ciolek98}; Ogino, Tomisaka, \& Nakamura \cite{Ogino}).
It is always present since the outer boundary condition for collapse
is distinct from the inner limit of self-similar supersonic
infall found in the Larson-Penston solution. Rather, the outer boundary
condition must represent 
the ambient conditions of a molecular cloud, which do not correspond 
to large-scale infall (Zuckerman \& Evans \cite{Zuck}).
Additionally, the finite mass reservoir and steeper than $r^{-3}$ profile
as a source of the declining accretion rate has been studied
analytically by Henriksen, Andr\'e, \& Bontemps (\cite{Henriksen}) and
Whitworth \& Ward-Thompson (\cite{Whit2}), although they did not account
for the physical origin of such a steep density slope.

Indeed, a study of outflow activity from young stellar objects (YSO's)
by Bontemps et al. (\cite{Bontemps}; hereafter BATC) suggests
that $\dot{M}$ declines significantly with time during the
accretion phase of protostellar evolution.
Specifically, BATC have shown that if the CO outflow rate is proportional
to $\dot{M}$, then Class~0 objects
(young protostars at the beginning of the main accretion phase) have
an $\dot{M}$ that is
factor of 10 greater (on average) than that of the more evolved Class~I 
objects.
In this paper, we investigate in detail how the assumption
of constant mass and volume of a gravitationally
contracting core can affect the mass accretion rate and other observable
properties after the formation of the central hydrostatic stellar core.
A very important question is: which of the two effects mentioned above -
a gradient of infall speed in the prestellar phase, or a finite mass
reservoir and associated steep outer density slope - is more relevant
to explaining the observations of BATC? The evolutionary tracks of
envelope mass $\Menv$ versus bolometric luminosity $\Lbol$ 
are another important diagnostic of protostellar evolution
(Andr\'e et al. \cite{Andre2}). BATC have fit the data using a toy model in which 
$\dot{M}$ decreases with time in exact proportion to the remaining envelope
mass $M_{\rm env}$, i.e. $\dot{M}=M_{\rm env}/\tau$, where $\tau$
is a characteristic time.

We seek to explain the observed YSO evolutionary tracks using a physical 
(albeit highly simplified) model. We perform high resolution one-dimensional 
spherical isothermal simulations.
The initial peak and decline in the
mass accretion rate is modeled through numerical simulations and a simplified 
semi-analytic approach. A second late-time decline in $\dot{M}$ due to a gas 
rarefaction wave propagating inward from the outer edge of a 
contracting core, is also studied
in detail. Comparisons are made with the observationally inferred
decrease of mass accretion rate (BATC), and evolutionary tracks of 
$\Menv$ versus bolometric luminosity $\Lbol$ (from Motte \& Andr\'e
\cite{Motte}).

Numerical simulations of spherical collapse of
isothermal cloud cores are described in \S~\ref{IC}.
The comparison of the model 
with observations is given in
\S~\ref{Obs}.  Our main conclusions are summarized in \S~\ref{Sum}.
An analytical approach for the determination of the
mass accretion rate is presented in the Appendix.

\section{Isothermal collapse}
\label{IC}

\subsection{Model Assumptions}
\label{MA}

We consider the gravitational collapse of
spherical isothermal (temperature $T=10$~K) 
clouds composed of molecular
hydrogen with a $10\%$ admixture of atomic helium.
The models actually represent cloud cores which are embedded within a 
larger molecular cloud.
The evolution is calculated by solving the
hydrodynamic equations in spherical coordinates:
\begin{eqnarray}
\frac{\partial \rho}{\partial t} + \frac{1}{r^2} \frac{\partial}{\partial r} \left(r^2 \rho v_r\right)  & = & 0\\
\frac{\partial}{\partial t} \left( \rho v_r \right) +
\frac{1}{r^2} \frac{\partial}{\partial r} \left(r^2 \rho v_r v_r\right) & = &
- \frac{\partial p}{\partial r} - \rho \frac{G\, M}{r^2}, \\
\frac{\partial e}{\partial t} + {1\over r^2}\frac{\partial}{\partial r}(r^2
e v_r) & = & - \frac{p}{r^2} \frac{\partial}{\partial r}(r^2 v_r)
\label{hydro}
\end{eqnarray}
where $\rho$ is the density, $v_r$ is the radial velocity, $M$ is the
enclosed mass, $e$ is the internal energy density and $p=(\gamma-1) e$
is the gas pressure. The ratio of specific heats is equal to $\gamma=1.001$
for the gas number density $n\le10^{11}$~cm$^{-3}$, which implies isothermality
(the value of $\gamma$ is not exactly unity in our implementation in
order to avoid a division by zero). We define the gas number density $n=\rho/m$,
where $m=2.33~m_{\rm H}$ is the mean molecular mass.
When the gas number density in the collapsing core
exceeds $10^{11}$~cm$^{-3}$, we form the central hydrostatic stellar
core by imposing an adiabatic index $\gamma=5/3$.
This simplified treatment of the transition to an opaque protostar
misses the details of the physics on small scales. Specifically,
a proper treatment of the accretion shock and radiative transfer effects is 
required to accurately predict the properties of the stellar
core (see Winkler \& Newman \cite{Winkler} for a detailed treatment and
review of work in this area). 
However, our method should be adequate to study the protostellar
accretion rate, and has been used successfully by e.g.
Foster \& Chevalier (\cite{FC}) and Ogino et al. (\cite{Ogino})
for this purpose.
We use the method of finite-differences, with the time-explicit,
operator split
solution procedure used in the ZEUS-1D numerical hydrodynamics code; it is
described in detail by Stone \& Norman (\cite{Stone}).
We have introduced the momentum density correction factor, as advocated
by M\"{o}nchmeyer \& M\"{u}ller (\cite{MM}), to avoid the development of
an anomalous density spike at the origin (see Vorobyov \& Tarafdar \cite{VT}
for details).
The numerical grid has 700 points which are initially uniformly spaced, 
but then move with the gas until the central
stellar core is formed. This provides an adequate resolution throughout
the simulations. 

We impose boundary conditions such that the gravitationally bound cloud
core has a constant mass and volume.
The assumption of a constant mass appears to be observationally
justified by the sharp outer density profiles described in \S\ \ref{Intro}.
Physically, this assumption may be justified if the
core decouples from the rest of a comparatively static, diffuse cloud
due to a shorter dynamical timescale
in the gravitationally contracting central condensation than in the
external region. A specific example of this, due to enhanced magnetic 
support in the outer envelope, is found in the models of ambipolar-diffusion 
induced core formation (see, e.g. Basu \& Mouschovias \cite{Basu95}).
The assumption of a constant volume is mainly an
assumption of a constant radius of gravitational influence of a cloud core 
within a larger parent diffuse cloud.

The radial gas density distribution of a self-gravitating cloud
with finite central density that is in hydrostatic equilibrium (e.g.
Chandrasekhar \cite{Chandra}) can be conveniently approximated by a modified
isothermal sphere, with gas density
\begin{equation}
\rho=\frac{ \rho_{\rm c}}{1+(r/r_{c})^2}
\label{modiso}
\end{equation}
(Binney \& Tremaine \cite{BT}), where
$\rho_{\rm c}$ is the central density and
$r_{\rm c}$ is a radial scale length. We choose a value
$r_{\rm c}=1.1~c_{s}/\sqrt{\pi G \rho_{\rm c}}$, so that the inner profile
is close to that of a Bonnor-Ebert sphere, $\rc$ is comparable
to the Jeans length, and the asymptotic density profile is 2.2 times the
equilibrium singular isothermal sphere value $\rho_{\rm SIS} =
\cs^2/(2 \pi G r^2$). The latter is justified on the grounds that 
core formation should occur in a somewhat non-equilibrium manner
(an extreme case is the Larson-Penston flow, in which case the
asymptotic density profile is as high as $4.4 \, \rho_{\rm SIS}$),
and also by observations of protostellar envelope density profiles
that are often overdense compared to $\rho_{\rm SIS}$ (Andr\'e,
Motte, \& Belloche \cite {Andre3}). 

For small radii ($r \leq \rc$), the initial density is very close
to the equilibrium
solution for an isothermal sphere with a finite central density.
However, at large radii it is twice the value
of the equilibrium isothermal sphere, which converges 
to $\rho_{\rm SIS}$. Hence, our initial conditions resemble
those of other workers 
(Foster \& Chevalier \cite{FC}; Ogino et al. \cite{Ogino})
who start with Bonnor-Ebert spheres and add a positive density 
perturbation in order to initiate evolution. Use of the modified isothermal
sphere simplifies the analysis a little bit since there is a clear
transition from flat central region to a power-law outer profile.
The choice of central density $\rho_{\rm c}$ and outer radius $\rout$ determines the
cloud mass. We study many different cloud masses - two models are
presented in this section and other models are used to fit observational
tracks in \S\ \ref{Obs}.
We also add a (small) positive density perturbation of
a factor of 1.1 (i.e. the initial gas density distribution is increased 
by a factor of 1.1) to drive the cloud (especially the inner region 
which is otherwise near-equilibrium) into gravitational collapse. 

Table~\ref{Table1} shows the
parameters for two model clouds presented in this section.
The adopted central number density $n_{\rm c}=5\times 10^{4}$~cm$^{-3}$
is roughly an order of magnitude lower than is observed in prestellar cores
(Ward-Thompson et al. \cite{WT}). Considering that these cores may be already in
the process of slow gravitational contraction, our choice of $n_{\rm c}$
is justified for the purpose of describing the basic features of 
star formation. 
In both models, the outer radius $r_{\rm out}$ is chosen so as
to form gravitationally unstable prestellar cores with central-to-surface
density ratio $\rho_{\rm c}/\rho_{\rm out}\ge 14$ (since our initial states
are similar to Bonnor-Ebert spheres). In model~I1, $\rho_{\rm c}/\rho_{\rm out}\approx 24$
and by implication $r_{\rm out}/r_{\rm c} \approx 4.7$, whereas in model~I2 
$\rho_{\rm c}/\rho_{\rm out}\approx 324$ and $r_{\rm out}/r_{\rm c} \approx 18$.
Model~I2 thus represents a very extended prestellar core.
Models~I1 and I2 have masses $5~M_{\odot}$ and $24~M_{\odot}$
respectively; the `I' stands for isothermal.

\begin{table}
\caption{Model parameters}
\label{Table1}
\vskip 0.1cm
\begin{tabular}{llllllll}
\hline
Model & $n_{\rm c}$ & $r_{\rm c}$ & $r_{\rm out}$ &${\rho_{\rm c}\over
\rho_{\rm out}}$ & $M_{\rm cl}$ & $T$  \\ [2 pt]
\hline
I1 & $5.0$ & 0.033 & 0.16 &24 & 5 & 10 \\
I2 & $5.0$ & 0.033 & 0.5 &324 &24 & 10  \\
\hline
\end{tabular}

\medskip
All number densities are in units of $10^4$~cm$^{-3}$, lengths are in pc,
masses in $M_\odot$, and temperatures in K.
\end{table}

\subsection{Numerical Results}
\label{RES}

\begin{figure}
\includegraphics[width=84mm]{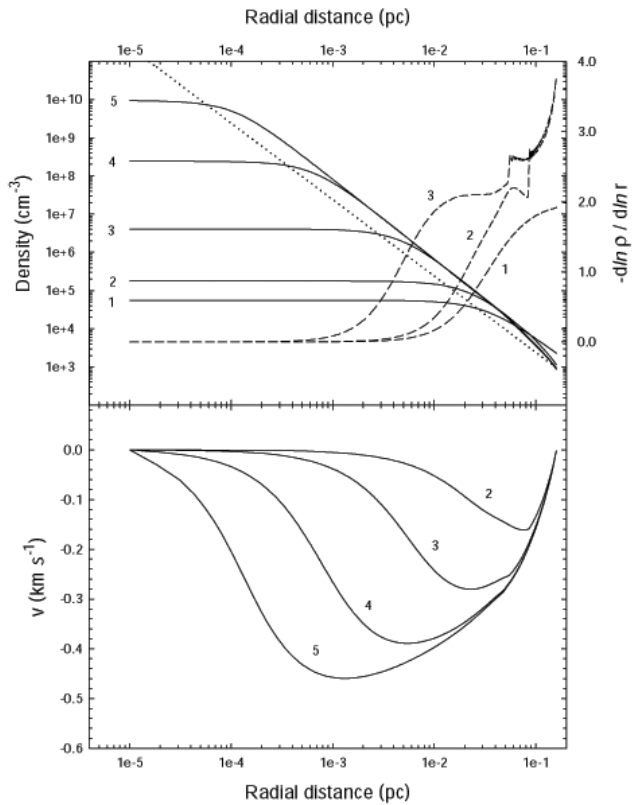}
      \caption{The radial gas density (the upper panel) and velocity (the
      lower panel) profiles obtained in model~I1 before the central hydrostatic
      stellar core is formed. The number 1 corresponds to the initial
      profiles (note
      that initially the cloud core is at rest) and the number 5 labels profiles
      when the central gas number density has almost reached $10^{10}$~cm$^{-3}$.
      The dashed lines in the upper panel
      show the power-law index  $d \ln \rho / d \ln r$
      of the gas distributions 1, 2, and 3. Profiles 2, 3, 4, and 5 are
      reached at times 0.309 Myr, 0.378 Myr, 0.392 Myr, and 0.394 Myr,
      respectively. For reference, the dotted line
      is the density profile of a singular isothermal sphere.}
         \label{fig1}
\end{figure}

Fig.~\ref{fig1} shows the temporal evolution of the radial gas
density profiles (the upper panel) and velocity profiles (the lower panel)
during the runaway collapse phase (before the
formation of the central hydrostatic stellar core) in model~I1.
The density and velocity
profiles are numbered according to evolutionary sequence, starting from
the initial distributions (profile~1; note that the cloud core is initially at
rest) and ending with those obtained when the central number density has
almost reached $10^{10}$~cm$^{-3}$ (profile~5).
The dashed lines in the upper panel of Fig.~\ref{fig1} show the
power-law index $d \ln \rho/d \ln r$
of the gas distribution for profiles 1, 2, and 3.
By the time that a relatively mild 
center-to-boundary density contrast $\sim 150$ is established (profile 2),
the radial density profile starts resembling those observed in Taurus by Bacmann
et al. (\cite{Bacmann}): it is flat in the central region,
then gradually changes to an $r^{-2}$ profile, and falls off as $r^{-3}$ 
or steeper in the envelope at $r\ga 0.08$~pc.
The sharp change in slope of the density profile (e.g. at $r \sim
0.08$~pc in profile 2 of Fig.~\ref{fig1}) {\it is due to an inwardly-propagating
gas rarefaction wave
caused by a finite reservoir of mass}.
The self-similar region with $r^{-2}$ density profile is 
of the Larson-Penston type, with density somewhat
greater than the equilibrium singular isothermal sphere value
($\cs^2/2 \pi G r^2$).
The velocity profiles in  Fig.~\ref{fig1} also show a distinct break at
the instantaneous location of the rarefaction wave.
Furthermore, the peak infall speed is clearly supersonic 
(since $\cs = 0.19$ km s$^{-1}$) by the time profile 4 is established,
again consistent with Larson-Penston type flow in the inner region. 

\begin{figure}
\includegraphics[width=84mm]{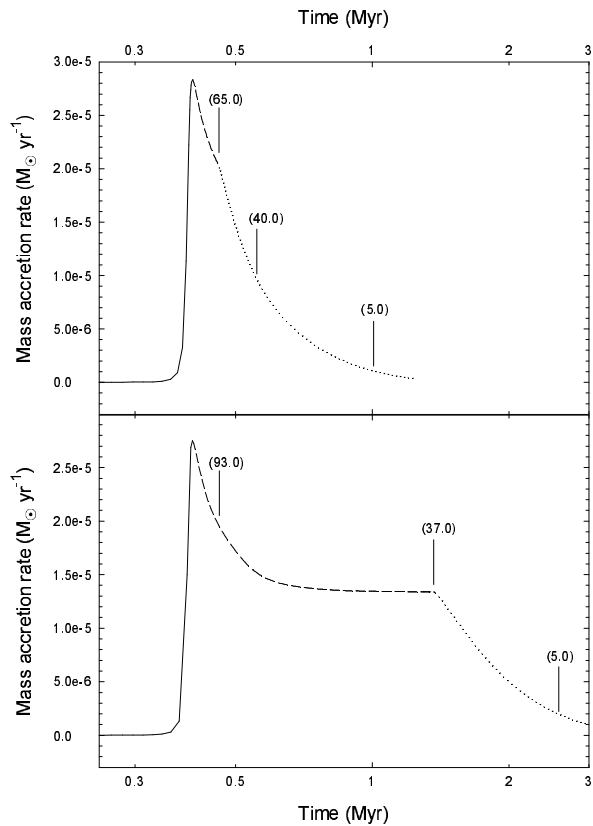}
      \caption{The temporal evolution of the mass accretion rate at the
      radial distance $r=600$~AU
      from the center of a cloud core obtained in {\bf a)} model~I1 and
      {\bf b)} model~I2. The model cloud I1 has mass 5 $\Msun$
      and the model cloud I2 has mass 24 $\Msun$.
      The solid lines show $\dot{M}$ during the runaway collapse phase,
      prior to the formation of a central stellar core.
      The dashed and dotted lines plot $\dot{M}$ after stellar
      core formation;
      the dashed lines show $\dot{M}$ before the gas affected by
      the inwardly propagating rarefaction
      wave has reached $r=600$~AU, whereas the dotted lines show $\dot{M}$
      after this gas has reached $r=600$~AU. 
        The numbers in parentheses reflect the percentage of cloud mass
        remaining in the envelope at the given times.}
         \label{fig2}
\end{figure}

Fig.~\ref{fig2}a shows the temporal evolution of the accretion rate
at a radial distance of 600~AU from the center in model~I1.\footnote{
We note that the accretion rate is not expected to vary significantly 
in the range $0.1~{\rm AU} <r<1000$~AU according to Masunaga \& Inutsuka
(\cite{MI}).} The evolution 
is characterized by a slow initial gravitational contraction
and then a very rapid increase until about 0.4~Myr. Subsequently, a central
hydrostatic stellar core forms and the mass accretion
rate reaches its maximum value of $2.8 \times 10^{-5}~M_{\odot}$~yr$^{-1}$
(or $17.4~c_{\rm s}^3/G$). 
After stellar core formation, the evolution of the mass
accretion rate has possibly {\it three distinct phases}, of which
two are on display in  Fig.~\ref{fig2}a. The {\it early phase}, plotted
with the dashed line in Fig.~\ref{fig2}a,
is characterized by accretion of material that has not yet been affected
by the rarefaction wave propagating inward from the outer boundary.
The accretion rate is declining, even though the density profile near the
stellar core was nearly self-similar at the moment of its formation.
This decline is due to the gradient of infall velocity in the inner regions,
an effect not predicted in the similarity solutions. However, if there is
a large outer region with mass
shells that are falling in at significantly subsonic speeds when the central 
stellar core forms (see discussion of Fig. ~\ref{fig2}b below),
the accretion
rate will eventually stabilize to a constant value that is consistent with
the standard theory of Shu (\cite{Shu}). In that picture,
progressively higher shells of gas lose their partial pressure support
and start falling from rest on to the central stellar core almost in a
free-fall manner. This would be the {\it intermediate phase} of accretion.
However, the {\it late phase} of very rapid decline of the accretion rate starts
at roughly 0.46~Myr (before the intermediate phase can be established
in the $5 \, \Msun$ cloud),
when gas affected by the inwardly propagating rarefaction wave reaches
the inner 600~AU. This results in a sharp drop of $\dot{M}$
as shown in Fig.~\ref{fig2}a by the dotted line.

The existence of the (in principle) three distinct phases of mass accretion
is clearly seen Fig.~\ref{fig2}b, where $\dot{M}$
of the more extended cloud ($r_{\rm out}/r_{\rm c} \approx 18$)
is plotted (hereafter, model I2).
The outer boundary is now at $r_{\rm out}=0.5$~pc and it takes a 
time $\ga 1$~Myr for the influence of the rarefaction wave to reach
the inner 600~AU.
As a result, the mass accretion rate
has time to stabilize at a constant value of $1.34\times
10^{-5}~M_\odot$~yr$^{-1}$
(the dashed line in Fig.~\ref{fig2}b), before it sharply drops at later times
(the dotted line in Fig.~\ref{fig2}b).
According to Shu (\cite{Shu}), the collapse from rest of a power-law
profile that has a density equal to twice $\rho_{\rm SIS}$ 
yields a mass accretion rate
$5.58 \, \cs^3/G = 8.86 \times 10^{-6}~M_\odot$~yr$^{-1}$. 
Our stable intermediate accretion
rate is roughly consistent with this prediction since the density in the
power-law tail is actually somewhat greater than twice $\rho_{\rm SIS}$.
It is equal to 2.42~$\rho_{\rm SIS}$ in the initial state, and grows
to greater overdensities in the innermost regions. However, the
bulk of the matter, which is in the outer tail, has density within 
$2.5 \rho_{\rm SIS}$.
Further experiments with our numerical simulations show that the intermediate
phase of constant accretion rate is observed only in rather extended
prestellar cores with $r_{\rm out}/r_{\rm c} \ga 15$. 
Foster \& Chevalier (\cite{FC}) found an even stronger criterion 
$r_{\rm out}/r_{\rm c} \ga 20$.
Since more extended cores tend to be
more massive as well, we may expect to observe the intermediate phase 
more frequently in the collapse of massive cores.

\subsection{Effect of Boundary Condition}
\label{BC}

Our standard simulation does not contain an external medium
explicitly. In order to explore the effect of such a medium, 
we ran additional simulations in which the cloud core is surrounded
by a spherical shell of diffuse (i.e.
non-gravitating) gas of constant temperature and density. 
The outermost layer of the cloud core and the external gas are initially 
in pressure balance.
We found that the value of $\dot{M}$ in the late accretion phase
may depend on the assumed values of the external density and temperature.
For instance, if the gravitating core is nested within a larger diffuse 
non-gravitating cloud of $T=10$~K
and $\rho=\rho_{\rm out}$, the accretion rate increases slightly as compared to that shown in Fig.~\ref{fig2} by the dotted line.
A warmer external non-gravitating environment
of $T=200$~K and $\rho=\rho_{\rm out}/20$ shortens the duration of the late
accretion phase shown in Fig.~\ref{fig2} by the dotted line.
This phase may be virtually absent if the sound speed of the external
diffuse medium is considerably higher (by a factor $\sim 1000$) than that
of the gravitationally bound core.
This essentially corresponds to a constant outer pressure boundary condition
(see Foster \& Chevalier \cite{FC}).
However, such a high sound speed contrast is not expected
for star formation taking place
in a dense ($n\sim 10^{4}$~cm$^{-3}$) environment like $\rho$ Ophiuchi
(Johnstone et al. \cite{Johnstone}).

We believe that the constant volume boundary condition, and resulting
inward propagating rarefaction wave, are best at reproducing the 
steep outer density profiles and the low (residual) mass accretion rate
necessary to explain the Class~I phase of protostellar accretion.

\subsection{Semi-analytic Model}
\label{SEMIANAL}

\begin{figure}
\includegraphics[width=84mm]{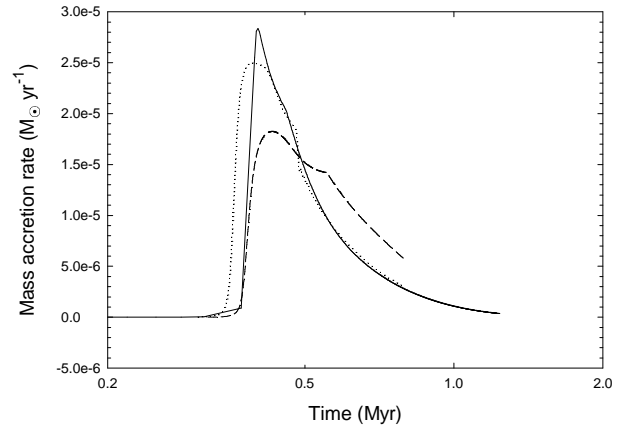}
      \caption{The temporal evolution of the mass accretion rate.
      The solid line shows the results of isothermal numerical simulations
      (model~I1). The
      dashed line shows $\dot{M}$ obtained in a pressure-free approximation
      if such collapse begins from rest with the relatively mildly
      concentrated density profile 2 in Fig.~\ref{fig1}.
      The dotted line shows the result for pressure-free collapse if
      a non-zero initial velocity (that of profile 2 in the bottom panel of
      Fig.~\ref{fig1}) is also used. The agreement of the pressure-free
      model and full numerical simulation are quite good in the latter
      case.}
         \label{fig3}
\end{figure}

Finally, we compute $\dot{M}$ of a pressure-free cloud using
the analytical approach developed in the Appendix.
This approach allows for the determination of $\dot{M}$ for a
cloud with given initial radial density  $\rho(r_{\rm 0})$
and velocity $v_{\rm 0}(r_{\rm 0})$ profiles, if the subsequent collapse
is pressure-free.
We find that the success or failure of the analytical approach to describe
the mass accretion rate of the isothermal cloud depends on the adopted
$\rho(r_{\rm 0})$ and $v_{\rm 0}(r_{\rm 0})$ profiles.
For instance, if $\rho(r_{\rm 0})$ is determined by profile 2
(the upper panel of Fig.~\ref{fig1}) and $v_{\rm 0}(r_{\rm 0})=0$,
the pressure-free mass accretion rate shown in Fig.~\ref{fig3}
by the dashed line
reproduces only very roughly the main features of the isothermal accretion rate
(the solid line in Fig.~\ref{fig3}).
However, if we take into account the non-zero and non-uniform velocity profile
$v_{\rm 0}(r_{\rm 0})$ plotted in the lower panel of
Fig.~\ref{fig1} (profile~2),
then the pressure-free $\dot{M}$ shown by the dotted line in Fig.~\ref{fig3}
reproduces that of the isothermal cloud much better.
This example demonstrates the importance of the velocity field {\it prior}
to stellar core formation in
determining the accretion rates {\it after} its formation.
The success of our
analytical pressure-free approach also shows that
the collapse of the isothermal cloud can be regarded as essentially
pressure-free from the time of a relatively mild central concentration
$\rho_{\rm c}/\rho_{\rm out} \ga 150$, when the central number density
$\sim 2 \times 10^5$ cm$^{-3}$.

\section{Astrophysical implications}
\label{Obs}

\begin{figure}
\includegraphics[width=84mm]{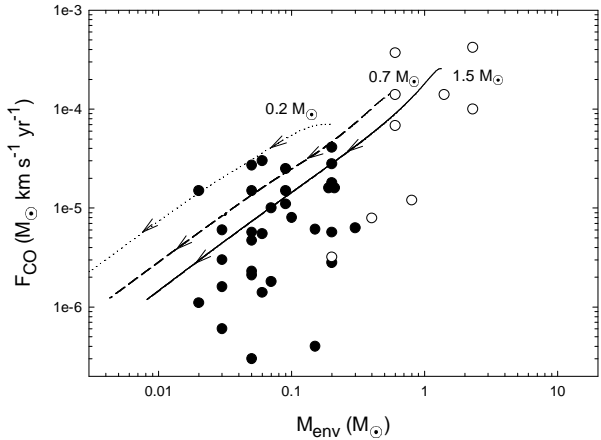}
      \caption{CO momentum flux $F_{\rm co}$ versus envelope mass
      $\Menv$ for 41 sources 
      of Bontemps et al. (1996). 
        The Class~0 and Class~I objects are plotted
      with the open and filled circles, respectively. The model $\Fco
      -\Menv$ tracks of three prestellar clouds of $M_{\rm cl}=0.2~\Msun$,
      $0.7~\Msun$, and $2~M_\odot$ are shown by the dotted, dashed, and  solid lines, respectively.} 
         \label{fig4}
\end{figure}

Class~0 objects represent a very early phase of protostellar
evolution (see Andr\'e et al. \cite{Andre2}), 
as evidenced by a relatively high ratio of 
submillimeter luminosity to bolometric luminosity: 
$L_{\rm submm}/\Lbol > 0.5\%$. 
Class~0 objects also drive powerful collimated CO outflows.
A study of outflow activity in low-mass YSO's by
BATC suggests that the CO momentum flux $F_{\rm
co}$ declines significantly during protostellar evolution. Specifically,
$F_{\rm co}$ decreases on average by more than an order of magnitude 
from Class~0 to Class~I objects.
This tendency is illustrated in Fig.~\ref{fig4}, where we plot
$\Fco$ versus $\Menv$ for 41 sources listed in BATC. 
We relate $\Fco$ to $\dot{M}$ by
\begin{equation}
F_{\rm co}=f_{\rm ent} \times (\dot{M}_{\rm w}/\dot{M}) V_{\rm w} \times
\dot{M},
\label{CO}
\end{equation}
where $f_{\rm ent}$ is the entrainment efficiency that relates $F_{\rm co}$
to the momentum flux $\dot{M}_{\rm w} V_{\rm w}$ of the wind.
Based on theoretical models in the literature, BATC suggested that the 
factor $f_{\rm ent}$ and the outflow
driving engine efficiency $(\dot{M}_{\rm w}/\dot{M}) V_{\rm w}$ 
do not vary significantly during protostellar
evolution. This implies that the observed decline of $\Fco$ reflects 
a corresponding decrease in $\dot{M}$ from the Class~0 to
the Class~I stage.
Following BATC, we take $f_{\rm ent}=1$, $\dot{M}_{\rm w}/\dot{M}=0.1$, and
$V_{\rm w}=150$~km~s$^{-1}$ and use equation (\ref{CO}) to compute 
$\Fco$ from our model's known mass accretion rate $\dot{M}$
(see Fig.~\ref{fig2}). 

Since the sample of Class 0 and Class I objects listed in BATC includes 
sources from both the Ophiuchus and Taurus star forming regions, we develop
model clouds which take into account the seemingly different initial conditions
of star formation in these regions.
As mentioned in \S~\ref{Intro}, the two most prominent differences between
these two regions are: 
(1) The cores in Ophiuchus have 
outer radii ($r_{\rm out} \la 5000$~AU) which are smaller than in Taurus, where
$5000~{\rm AU}~\la~r_{\rm out}~\la~20000$~AU  
(Andr\'e et al. \cite{Andre}; Andr\'e et al. \cite{Andre2});
(2) The radial column density profiles
of the protostellar envelopes of Class~0 objects in Ophiuchus
are at least 2-3 times denser than a SIS at $T=10$~K, whereas in Taurus the protostellar
envelopes are overdense compared to the SIS by a smaller factor $\la 2$ (Andr\'{e} et al.
\cite{Andre3}).
This implies that radial column density profiles of {\it prestellar} cores 
in Ophiuchus and Taurus may follow the same tendency. 
We develop a set of Ophiuchus model cores which have $r_{\rm out}\la 5000$~AU,
and a set of Taurus model cores which have $5000~{\rm AU}~\la~r_{\rm out}~\la~20000$~AU.
Furthermore, the factor $\alpha$ (by which our model density profiles
are asymptotically overdense compared to $\rho_{\rm SIS}$) is taken to
be $\ga 2.0$ for Ophiuchus and $<2.0$ for Taurus.  
Clearly, there is no unique set of model cloud parameters that would be exclusively
consistent with the observational data, given the measurement uncertainties. 
We have chosen a set of core central densities $\rhoc$, radii $\rout$, and overdensity factors
$\alpha$ so as to reasonably
reproduce the observed properties of the cores in the two regions.
The parameters of the model 
density distributions for Ophiuchus and Taurus are listed in Table~\ref{Table2}
and Table~\ref{Table3}, respectively.

We also note that we have ensured that the cores satisfy the gravitational 
instability criterion $r_{\rm out}/r_{\rm c}\ga 3.6$, which is similar to that for 
Bonnor-Ebert spheres. 
The Ophiuchus model cores are clustered near this limiting
value of $r_{\rm out}/r_{\rm c}$, but the Taurus model cores are 
allowed to be somewhat more extended, again in keeping with observed 
properties. 
We also note that the masses of prestellar cores with the radial density
profile given by equation (\ref{modiso}) scale as $1/\rho_{\rm c}^{0.5}$, if 
the ratio $r_{\rm out}/r_{\rm c}$ is fixed.

\begin{table}
\caption{Model parameters for Ophiuchus}
\label{Table2}
\vskip 0.1cm
\begin{tabular}{llllll}
\hline
$M_{\rm cl}$ & $n_{\rm c}$&  $r_{\rm out}$ &
$\rho_{\rm c}/\rho_{\rm out}$ & $r_{\rm out}/r_{\rm c}$ & $\alpha$ \\ [2 pt]
\hline
0.17 & $1\times 10^7$ & 1600 & 15.0 & 3.7 & 2.0 \\
0.23 & $5\times 10^6$ & 1900 & 14.1 & 3.6 & 2.0 \\
0.55 & $2\times 10^6$ & 4000 & 18.0 & 4.1 & 2.4 \\
0.9  & $2\times 10^6$ & 4000 & 18.0 & 4.1 & 4.0 \\
\hline
\end{tabular}

\medskip
All number densities are in cm$^{-3}$, lengths in AU, and masses
in $M_\odot$.
\end{table}

\begin{table}
\caption{Model parameters for Taurus}
\label{Table3}
\vskip 0.1cm
\begin{tabular}{llllll}
\hline
$M_{\rm cl}$ & $n_{\rm c}$&  $r_{\rm out}$ &
$\rho_{\rm c}/\rho_{\rm out}$ & $r_{\rm out}/r_{\rm c}$ & $\alpha$ \\ [2 pt]
\hline
0.46 & $1.5\times 10^6$ & 5000 & 18 & 4.1 & 1.9 \\
0.65 & $1\times 10^6$ & 6000 & 26 & 5.0 & 1.8 \\
1.0 & $1\times 10^6$ & 10000 & 71 & 8.4 & 1.5 \\
1.5 & $8\times 10^5$ & 12000 & 73 & 8.5 & 1.8 \\
\hline
\end{tabular}

\medskip
All number densities are in cm$^{-3}$, lengths in AU, and masses
in $M_\odot$.
\end{table}

The sample of 41 sources in BATC contains Class~0 and Class~I objects from
both Ophiuchus and Taurus. Hence, in Fig.~\ref{fig4} we take three representative 
prestellar clouds of $M_{\rm cl}=0.23~M_{\odot}$ (Ophiuchus),
$0.65~M_{\odot}$ (Taurus), and $2.0~M_{\odot}$ (Taurus),
for which the $F_{\rm co}-\Menv$ tracks
are shown by the dotted, dashed, and solid lines, respectively.
Both the data and model tracks show a near-linear correlation between 
$\Fco (\propto \dot{M})$ and $\Menv$. A slightly better fit of the model
tracks to the data can be obtained by adjusting one or more of the 
estimated parameters $f_{\rm ent}$, $\dot{M}_{\rm w}/\dot{M}$, 
and $V_{\rm w}$ by factors of order unity.

Based on the near-linear correlation of $F_{\rm co}$ and $\Menv$,
BATC developed a toy model in which
$\dot{M}$ decreases with time in exact proportion to the remaining envelope
mass $M_{\rm env}$, i.e. $\dot{M}=M_{\rm env}/\tau$, where $\tau$
is a characteristic time. Furthermore, if one assumes that the bolometric
luminosity derives entirely from the accretion on to the hydrostatic
stellar core,
i.e., $\Lbol=G M_{\rm c} \dot{M}/ R_{\rm c}$, where $M_{\rm c}$
and $R_{\rm c}$ are the mass and radius of the stellar core, respectively,
then the bolometric luminosity
reaches a maximum value when half of the initial prestellar mass has
been accreted by the protostar and the other half remains in the envelope. 
The evolutionary time when $M_{\rm c}=M_{\rm env}$
was defined by  Andr\'{e} et al. (\cite{Andre93}) as the conceptual border
between the Class~0 and Class~I evolutionary stages.

\begin{figure}
\includegraphics[width=84mm]{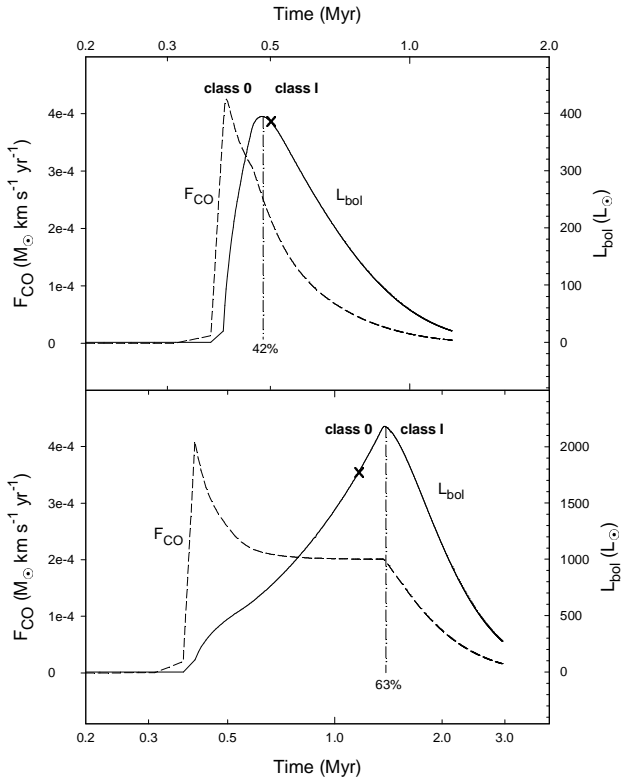}
      \caption{Temporal evolution of the bolometric luminosity $\Lbol$ and 
      CO momentum flux $F_{\rm co}$. {\bf a)} The solid and dashed lines 
      show $\Lbol$ and $F_{\rm co}$ obtained in model~I1, respectively.  
      {\bf b)} The solid and dashed lines show $\Lbol$ and $F_{\rm co}$ obtained 
      in model~I2, respectively. Note that $\Lbol$ is still increasing during the early phase 
      of accretion rate decline but only declines later due to the more severe 
      accretion rate decline caused by the inward propagating rarefaction wave.
      The vertical dash-dotted line is the temporal dividing line between the
      Class 0 and Class I phases for each model; the numbers below give
      the mass of the central stellar core as a percentage of the total cloud mass.
      Crosses indicate the time
      when 50\% of the initial cloud mass has been accreted by
      the protostar. Note the use of a logarithmic scale for time, 
      so that the Class~I
      phase is still longer than the Class~0 phase for model~I2.}
         \label{fig5}
\end{figure}

The solid and dashed lines in Fig.~\ref{fig5}a and Fig.~\ref{fig5}b
show $\Lbol$ and $F_{\rm co}$ obtained in model~I1 and model~I2,
respectively.
Since we do not follow the evolution of a
protostar to the formation of the second (atomic) hydrostatic core, we
take $R_{\rm c}= 3~R_\odot$ and let $\Lbol=G M_{\rm c} \dot{M}/ R_{\rm c}$.
The radius $ R_{\rm c}$ depends on the accretion
rate and stellar mass (see Fig.~7 of Stahler \cite{Stahler}) and may vary
from $R_{\rm c}\approx 1.5~R_{\odot}$ for small stellar cores $M_{\rm c}\sim 0.2~M_\odot$ 
and low accretion rates $\dot{M}\sim 2\times 10^{-6}~M_\odot$~yr$^{-1}$ to 
$R_{\rm c}\approx 5.0~R_{\odot}$ for large stellar cores $M_{\rm c}\sim 1.0~M_\odot$
and high accretion rates $\dot{M}\sim 1\times 10^{-5}~M_\odot$~yr$^{-1}$. 
However, this variation constitutes roughly a factor of 2 change in the adopted
average value of $R_{\rm c}= 3~R_\odot$. Indeed, we performed numerical
simulations with a varying $R_{\rm c}$ (assuming a normal deuterium abundance) 
and found that it has only a minor qualitative
influence on our main results. The stellar
core mass $M_{\rm c}$ is computed by summing up the masses of
the central hydrostatic spherical layers in our numerical simulations.
An obvious difference in the temporal evolution of $F_{\rm co}$ and $\Lbol$
is seen in Fig.~\ref{fig5}. The temporal evolution of $F_{\rm co}$
after the central hydrostatic core formation at $t\approx 0.4$~Myr
goes through the same phases as shown for $\dot{M}$ in Fig.~\ref{fig2}. 
The temporal evolution of $\Lbol$ shows {\it two} distinct phases:
it {\it increases during the early phase}
(unlike $F_{\rm co}$)
{\it and starts decreasing only when gas affected by the inward
propagating rarefaction wave reaches the central
hydrostatic core}. Thus, in our model, only the rarefaction wave
acts to reduce $\Lbol$ during the accretion phase of protostellar evolution.
This is a physical explanation for the peak in $\Lbol$ that also
occurs in the toy model of BATC. In that model, the bolometric luminosity 
reaches a maximum value when exactly half of the initial prestellar mass 
has been accreted by the protostar. In our simulations, the peak in $L_{\rm bol}$ 
corresponds to the evolutionary time
when $50\% \pm 10\% $ of the matter is in the protostar (higher
deviations up to  $+15\% $ are found in very massive and extended prestellar
clouds).

\begin{figure}
\includegraphics[width=84mm]{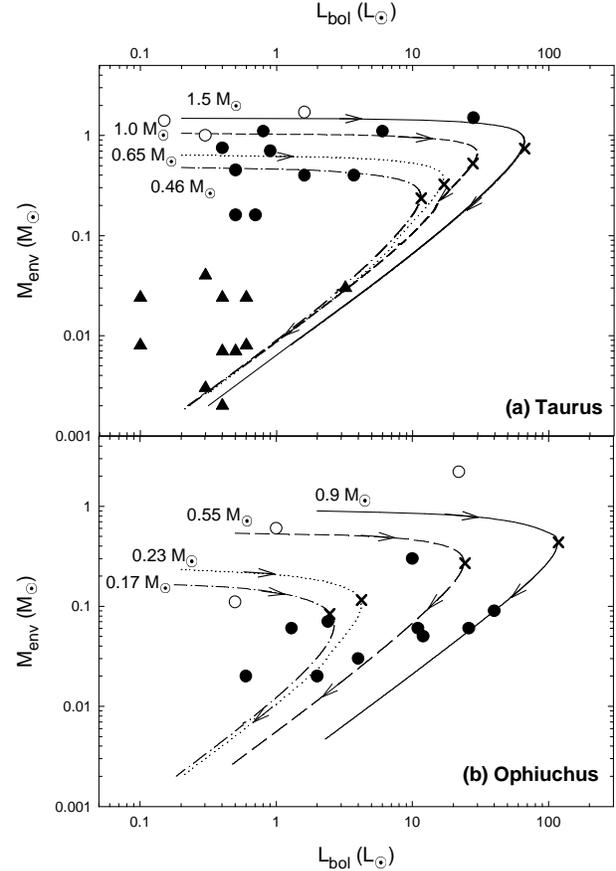}
      \caption{Envelope mass $M_{\rm env}$ versus bolometric luminosity
      $\Lbol$ for 37 protostellar objects taken from Motte \& Andr\'{e} (2001). 
      Diagrams are shown for a) Taurus, and b) Ophiuchus.
      The Class~0 and Class~I objects are plotted
      with the open and filled circles, respectively. The triangles represent
      the observed peculiar Class~I sources. The model $M_{\rm
      env}-\Lbol$ tracks of eight prestellar clouds with masses 
      $M_{\rm cl}=1.5~\Msun$,
      $1.0~\Msun$, $0.65~\Msun$, and $0.46~M_\odot$ (Taurus) and $0.9~\Msun$,
      $0.55~\Msun$, $0.23~\Msun$, and $0.17~\Msun$ (Ophiuchus)  are shown by 
      the solid, dashed, dotted, 
      and dotted-dashed lines, respectively. Crosses indicate the time when 
      $50\%$ of the cloud mass has been accreted by the protostar. }
         \label{fig6}
\end{figure}

Finally, in Fig.~\ref{fig6} we show the $M_{\rm env}-\Lbol$ evolutionary tracks. 
We use eight representative prestellar cloud core masses as listed in Table~\ref{Table2}
and Table~\ref{Table3}. Fig.~\ref{fig6}a shows the overlaid
data for YSO's in Taurus, while Fig.~\ref{fig6}b has overlaid data for 
Ophiuchus. 
The data for both samples are taken from Motte \& Andr\'e (\cite{Motte}).
The open circles represent bonafide Class~0 objects, the
solid circles represent the bonafide Class~I objects, while the 
triangles represent the so-called peculiar Class~I objects observed
in Taurus. 
We note that the envelope masses given in Table~2 of
Motte \& Andr\'e (\cite{Motte}) and plotted in their Fig.~5 and Fig.~6
are determined within a 4200~AU radius circle. While this should relatively
well describe {\it the total envelope masses} in Ophiuchus, a substantial
(a factor of 3) portion of the envelope mass may be missing in the
Taurus cores, which have sizes
as large as 15000-20000~AU. For this reason, we plot in Fig.~\ref{fig6}a the {\it total} 
envelope masses given in Table~4 of Motte \& Andr\'{e} (\cite{Motte}) for a set
of resolved Taurus cores.

The loci of maximum $\Lbol$ in the $\Menv-\Lbol$ tracks roughly
separate two phases in the evolution of a protostar: a shorter one characterized
by accretion of matter from the envelope
not yet affected by the rarefaction wave (i.e. characterized by the $r^{-2}$
gas density profile or shallower)
and a longer one characterized by accretion of
matter from the rarefied envelope (i.e. characterized by the $r^{-3}$ profile
or steeper). 
The turnover also corresponds to the evolutionary time
when $50\% \pm 10\%$ of the matter is in the protostar and a corresponding amount
remains in the envelope as shown by the crosses in Fig.~\ref{fig6}. 
This is in agreement with the observational 
requirements and toy model of BATC.
Given that the peak in $\Lbol$ is our conceptual dividing
line between two distinct phases of accretion, we conclude that
in Taurus, most of the so-called Class~I objects would tend to fall into the Class~0
category in our scheme. They may indeed be more evolved than the 
already identified Class~0 objects,
having lower values of $\dot{M}$ and $\Menv$, but would not be in a 
qualitatively distinct phase of evolution (see Motte \& Andr\'e \cite{Motte} for a similar
conclusion). 
In contrast, the so-called
peculiar Class~I objects in Taurus would be proper Class~I objects in
our scheme since they are likely in a phase of declining $\Lbol$.
In Ophiuchus, the currently identified Class~0 and 
Class~I objects do seem to fall on two distinct sides of the peak
in $\Lbol$.

Fig.~\ref{fig5} indicates that in extended clouds (as of model~I2)
the phase  of increasing $L_{\rm bol}$ is longer than in compact clouds (as of model~I1).
This may explain why this phase is more populated in Taurus
than in Ophiuchus. In addition, extended clouds have a longer phase
of accretion from the envelope not yet affected by the rarefaction wave
and, as a consequence, a higher probability of having a quasi-constant
accretion phase. This is in agreement with the previous suggestions made
by Henriksen et al. (\cite{Henriksen}) and Andr\'e et al. (\cite{Andre2}) that
the accretion history in Taurus is closer to the SIS scenario 
than in Ophiuchus. However, we note that none of our representative
prestellar cores listed in Table~\ref{Table3} and used to fit the
data for Taurus are extended enough  
($r_{\rm out}/r_{\rm c} \ga 15$) to have a distinct phase of 
constant accretion.

One problem should be pointed out here. While our model $L_{\rm bol}-M_{\rm env}$ 
tracks in Fig.~\ref{fig6} explain well the measured
bolometric luminosities in Ophiuchus, they
seem to overestimate $L_{\rm bol}$ in Taurus by a factor of 5-10. This
is the so-called ``luminosity problem'' that was first noticed by Kenyon, 
Calvet, \& Hartmann (\cite{Kenyon}). 
As a consequence, the position near the turnover in $L_{\rm bol}-M_{\rm
env}$  tracks for Taurus is scarcely populated.
This implies that while spherical collapse models may be appropriate
for the determination of $L_{\rm bol}$ in Ophiuchus, they tend
to overestimate $L_{\rm bol}$ in Taurus. 
It is possible that a significant magnetic regulation of the 
early stages of star formation in Taurus, as implied by e.g. 
polarization maps (Moneti et al. \cite{Moneti}) would yield
more flattened envelopes which result 
in a lower accretion rate on to the central protostar and a
smaller bolometric luminosity. Interestingly, Kenyon et al. (\cite{Kenyon}) 
also concluded that envelopes in Taurus should
be highly flattened in order to explain their spectral energy distribution.
Two dimensional simulations are required to address this issue.

Finally, it is worth noting that our Taurus model cores are generally
more massive than the Ophiuchus model cores. This in agreement with observations, and
can be justified theoretically on the basis of a lower mean column density 
(hence greater Jeans length and Jeans mass of a sheetlike configuration) in
Taurus compared to regions of more clustered star formation in e.g. Ophiuchus
and Orion. Taken at face value, our models 
then imply that Taurus protostars should be more massive in
general than Ophiuchus protostars. While such a conclusion must be tempered by the fact 
that we do not model magnetic support or feedback from outflows, there is 
some evidence that Taurus does have a significantly higher mass peak in its initial
mass function than does the Trapezium cluster in Orion
(see Luhman \cite{Luhman} and references within). 

\section{Conclusions}
\label{Sum}
Our numerical simulations indicate that the assumption of a finite
mass reservoir of prestellar cores is required to explain the observed
Class~0 to Class~I transition.
We start our collapse calculations by perturbing a modified isothermal sphere
profile (eq. [\ref{modiso}]) that is truncated and resembles 
a bounded isothermal equilibrium state.
Specifically, we find that

$\bullet$ Starting in the prestellar runaway collapse phase, a shortage
of matter developing at the outer edge of a core
generates an inward propagating
rarefaction wave that steepens the radial gas density profile in the envelope
from $r^{-2}$ to $r^{-3}$ or even steeper.

$\bullet$ After a central hydrostatic stellar core
has formed, and the cloud core has entered the accretion phase, the mass
accretion rate $\dot{M}$ on to the central protostar can be divided into three
possible distinct phases. In the early phase, $\dot{M}$ decreases due to
a gradient of infall speed that developed during the runaway collapse
phase (such a gradient is not predicted in isothermal similarity solutions).
An intermediate phase of near-constant $\dot{M}$ follows if the core
is large enough to have an extended zone of self-similar
density profile with relatively low infall speed during the 
prestellar phase. Finally, when accretion occurs from the region affected
by the inward propagating rarefaction wave, a terminal and rapid
decline of $\dot{M}$ occurs.

$\bullet$ A pressure-free analytic formalism for the mass accretion rate
can be used to predict the mass accretion rate after stellar core formation,
given the density and velocity profiles in a suitably late part of the
runaway collapse phase.
Our formulas can estimate $\dot{M}$ at {\it essentially any} radial distance
from the central singularity.
This makes it possible to obtain $\dot{M}$ as a function of radial
distance at any given time.
We have demonstrated the importance of the velocity field of a collapsing
cloud in determining $\dot{M}$; our approach successfully estimates
the accretion rate if the velocity field is taken into account.
It demonstrates that the initial decline in $\dot{M}$ is
due to the gradient of infall speed in the prestellar phase.

$\bullet$ From an observational point of view, we can understand evolutionary 
$\Menv-\Lbol$ tracks using core models of relatively small
mass and size, so that there is not an extensive self-similar
region, in agreement with the profiles observed by
e.g. Bacmann et al. (\cite{Bacmann}). This means that in the accretion
phase, $\dot{M}$ makes a direct transition from the early decline phase
to the late decline phase when matter is accreted from the region of
steep ($r^{-3}$ or steeper) density profile that is affected by the
inward propagating rarefaction wave.
In the first phase (which we identify as the true Class~0 phase), 
the bolometric luminosity $\Lbol$ is increasing with time, even though 
$\dot{M}$ and the CO momentum flux $F_{\rm co}$ are slowly decreasing.
In the second phase (which we identify as the Class~I phase), 
both $\Lbol$ and $F_{\rm co}$ decline with time.
Hence, our simulations imply that the influence of the rarefaction
wave roughly traces the conceptual border between the Class~0 and
Class~I evolutionary stages. Regions of star formation with
more extended cores, like Taurus, should reveal a larger fraction of protostars
in the phase of increasing $L_{\rm bol}$. Our Fig.~\ref{fig6} reveals that
this is indeed the case, if most of the so-called Class~I objects in Taurus are reclassified
as Class~0, according to our definition. The so-called peculiar Class~I
objects in Taurus would be bona-fide Class~I objects  according to our
definition (see Motte \& Andr\'e \cite{Motte} for a similar conclusion
on empirical grounds). 

$\bullet$ Luminosities derived entirely from the accretion on to the hydrostatic
stellar core tend to be larger than the measured bolometric luminosities
$L_{\rm bol}$ in Taurus by a factor of 5-10, while they seem to better explain
the measured $L_{\rm bol}$ in Ophiuchus. This implies that
physical conditions in Ophiuchus may favour a more spherically symmetric
star formation scenario.

Our results should be interpreted in the context of models of 
one-dimensional radial infall. They illuminate phenomena
which are not included in standard self-similar models of isothermal
spherical collapse, by clarifying the importance of boundary (edge)
effects in explaining the observed $\Fco-\Menv$ and $\Menv-\Lbol$ tracks.
Important theoretical questions remain to be answered, such as the nature
of the global dynamics of a cloud which could maintain a finite mass
reservoir for a core. 
A transition to a magnetically subcritical envelope may provide the 
physical boundary that we approximate in our model. For example,
Shu, Li, \& Allen (2004) have recently calculated the (declining) accretion 
rate from a subcritical envelope on to a protostar, under the assumption
of flux freezing. An alternate or complementary mechanism of limiting 
the available mass reservoir is the effect of protostellar outflows.

Our main observational inference is that a finite mass reservoir and
the resulting phase of residual accretion is
necessary to understand the Class~I phase of protostellar evolution.
Our calculated mass accretion rates really represent the infall onto an
inner circumstellar disk that would be formed due to rotation. 
Hence, our results are relatable to observations if matter is cycled through
a circumstellar disk and on to a protostar rapidly enough so that the 
protostellar accretion is at least proportional to the mass infall rate
on to the disk. This is likely, since disk masses are not observed to
be greater than protostellar masses, but needs to be addressed with a 
more complete model.
In future papers, we will investigate the role of non-isothermality
(using detailed cooling rates due to gas and dust),
rotation, magnetic fields, and non-axisymmetry in determining
$\dot{M}$ and implied observable quantities.

\section*{Acknowledgments}
We thank Sylvain Bontemps, the referee, for an insightful report which
led us to make significant improvements to the paper.
We also thank Philippe Andr\'e for valuable comments about the observational
interpretation of our results.
This work was conducted while EIV was supported by the NATO Science Fellowship Program
administered by the Natural Sciences and Engineering Research Council
(NSERC) of Canada. EIV also gratefully acknowledges present support
from a CITA National Fellowship.
SB was supported by a research grant from NSERC.

\appendix
\section{Pressure-Free Collapse}
\subsection{Collapse from rest}
The equation of motion of a pressure-free, self-gravitating spherically
symmetric cloud is
\begin{equation}
 {d v \over d t}= -{G M(r) \over r^2},
\label{motion}
\end{equation}
where $v$ is the velocity of a thin spherical shell at a radial
distance $r$ from the center of a cloud, and $M(r)$ is the mass inside
a sphere of radius $r$.
equation (\ref{motion}) can be integrated to yield the expression for velocity $v$
\begin{equation}
v={d r \over d t }=-\sqrt{2 G M(r_{\rm 0}) \left[{1\over r}-{1\over r_{\rm 0}}\right] },
\label{vel}
\end{equation}
where $r_{\rm 0}$ is the initial position of a mass shell at $t=0$, and
$M(r_{\rm 0})$ is the mass inside $r_{\rm 0}$.
Here, it is assumed that all shells are initially at rest: $v_{\rm 0}(r_{\rm 0})=0$.
Equation~(\ref{vel}) can be integrated by means of the
substitution $r=r_{\rm 0}\cos^2\beta$ (see Hunter \cite{Hunter}) to determine the time it takes
for a shell located initially at $r_{\rm 0}$ to move to a smaller radial distance $r$
due to the gravitational pull of the mass $M(r_{\rm 0})$. The answer is
\begin{equation}
t={\arccos{\sqrt{r/r_{\rm 0}}} +0.5 \sin(2 \arccos{\sqrt{r/r_{\rm
0}}}) \over \sqrt{2 G M(r_{\rm 0})/r^3_{\rm 0}}}.
\label{time}
\end{equation}
The velocity  $v(r,t)$ at a given radial distance $r$ and time $t$ can now be obtained
from equations (\ref{vel}) and (\ref{time}). The values of $r$ and $t$ are
sufficient to determine $r_0$ (a value $> r$ but $\leq r_{\rm out}$, where
$r_{\rm out}$ is the radius of a cloud) from equation (\ref{time}).
Subsequently, we use the obtained value of $r_0$ in equation (\ref{vel})
to obtain $v(r,t)$.

Provided that the shells do not pass through each other (i.e. the mass
of a moving shell is conserved, $dM(r,t)=dM(r_0,t_0)$),
the gas density of a collapsing cloud is
\begin{equation}
\rho(r,t)={\rho_{\rm 0}(r_{\rm 0}) \:r^2_{\rm 0}\:dr_{\rm 0} \over r^2\: dr},
\label{dens}
\end{equation}
where $\rho_{\rm 0}(r_{\rm 0})$ is the initial gas density at $r_{\rm
0}$. The ratio of $dr_{\rm 0}/dr$ determines how the thickness of a given
shell evolves with time. The relative thickness $dr_{\rm 0}/dr$ is
determined by differentiating  $r=r_{\rm 0}\cos^2\beta$ with respect to
$r_{\rm 0}$, yielding
\begin{equation}
{d r \over d r_{\rm 0}} = {r \over r_{\rm 0}} - r_{\rm 0} \sin {\left(2 \arccos\sqrt{r
\over r_{\rm 0}} \right)} \:
{d \beta \over d r_{\rm 0}}.
\label{drdr0}
\end{equation}
Next, $d \beta / d r_{\rm 0}$ is determined from an alternate form of
equation (\ref{time}):
\begin{equation}
\beta+0.5 \sin{2 \beta}=t \sqrt{\frac{2 G M(r_0)}{r_0^3}}.
\end{equation}
Differentiating with respect to $r_0$ yields
\begin{equation}
{d \beta \over d r_{\rm 0}} = \sqrt{ \frac{G}{2M(r_0)r_0^3}}
\left( \frac{dM(r_0)}{dr_0} - \frac{3M(r_0)}{r_0} \right).
\label{dbdr0}
\end{equation}

Now that the density $\rho(r,t)$ and velocity $v(r,t)$ distributions of
a collapsing pressure-free sphere are explicitly determined,
the mass accretion rate at any given radial distance $r$ and time
$t$ can be found as $\dot{M}(r,t)=4 \pi r^2 \rho(r,t) v(r,t)$.

\subsection{Collapse with non-zero initial velocity}
In a general case of non-zero initial radial velocity profile $v_{\rm 0}(r_{\rm
0})$,
integration of eqaution (\ref{motion}) yields
\begin{equation}
v={d r \over d t }=-\sqrt{2 G M(r_{\rm 0}) \left[{1\over r}-{1\over r_{\rm 0}}\right]+v_{\rm
0}^2(r_{\rm 0}) },
\label{vel2}
\end{equation}
where $v_{\rm 0}(r_{\rm 0})$ is the initial velocity of a shell at $r_{\rm 0}$.
Equation~(\ref{vel2}) can be reduced to an integrable
one by means of a substitution $r=r_{\rm 0} \cos^2\beta$
\begin{equation}
{\sin(2\beta)\: d\beta \over \sqrt{\tan^2{\beta}+a} }= dt \sqrt{{2 G M(r_{\rm
0}) \over r_{\rm 0}^3}},
\end{equation}
where $a=r_{\rm 0} v_{\rm 0}(r_{\rm 0}) / 2 G M(r_{\rm 0})$.
Another substitution $\sin^2{\beta}=x$ and integration over $x$ from $x=0$ to $x=\sin^2{\beta}$
finally gives the time $t$ it would take for a shell
located initially at $r_{\rm 0}$ and having a non-zero initial velocity
$v_{\rm 0}$ to move to a smaller radial distance $r$:
\begin{eqnarray}
&t&=\left({(1-a)2 G M(r_{\rm 0}) \over r_{\rm 0}^3}\right)^{-{1\over 2}} \:
\left( \sqrt{\left(1-{r \over r_{\rm 0}}+\delta\right)\left({r \over r_{\rm 0}}\right)} \right.
 \nonumber \\
& &  - \sqrt{\delta} - (1+\delta)\tan^{-1}{\sqrt{\delta}}  \nonumber \\
& &\left. + (1+\delta)\tan^{-1}{\sqrt{1-{r/r_{\rm 0}}+\delta} \over
\sqrt{r/r_{\rm 0}}}   \right) ,
\label{time2}
\end{eqnarray}
where $\delta=a/(1-a)$.

In the case of a non-zero initial velocity profile, it is more complicated
to obtain a simple analogue to equations (\ref{drdr0})-(\ref{dbdr0}) and
explicitly determine a density distribution $\rho(r,t)$, as done in
the previous example.
Instead, we obtain the mass accretion rate by computing the mass
that passes the sphere of radius $r$ during time $\triangle t$, i.e.
\begin{equation}
\dot{M}(r,t)={M(r_{\rm 0}+\triangle r_0)-M(r_{\rm 0}) \over \triangle t},
\label{accret}
\end{equation}
where $M(r_{\rm 0})$ is the mass inside a sphere of radius $r_{\rm 0}$.
A time interval $\triangle t$ is the time that it takes for two adjacent
shells of radius $r_0$ and $r_0+\triangle r_0$ to move to the radial
distance $r$. The value of $\triangle t$ can be found by solving  
equation (\ref{time2})
for fixed values of $r_0$, $r_0+\triangle r_0$, and $r$.

\subsection{Applications}
As two examples, we consider two different initial gas density profiles
and determine the pressure-free mass accretion rate $\dot{M}(r,t)$ as a function of radial
distance $r$ and time $t$.

\begin{figure}
\includegraphics[width=84mm]{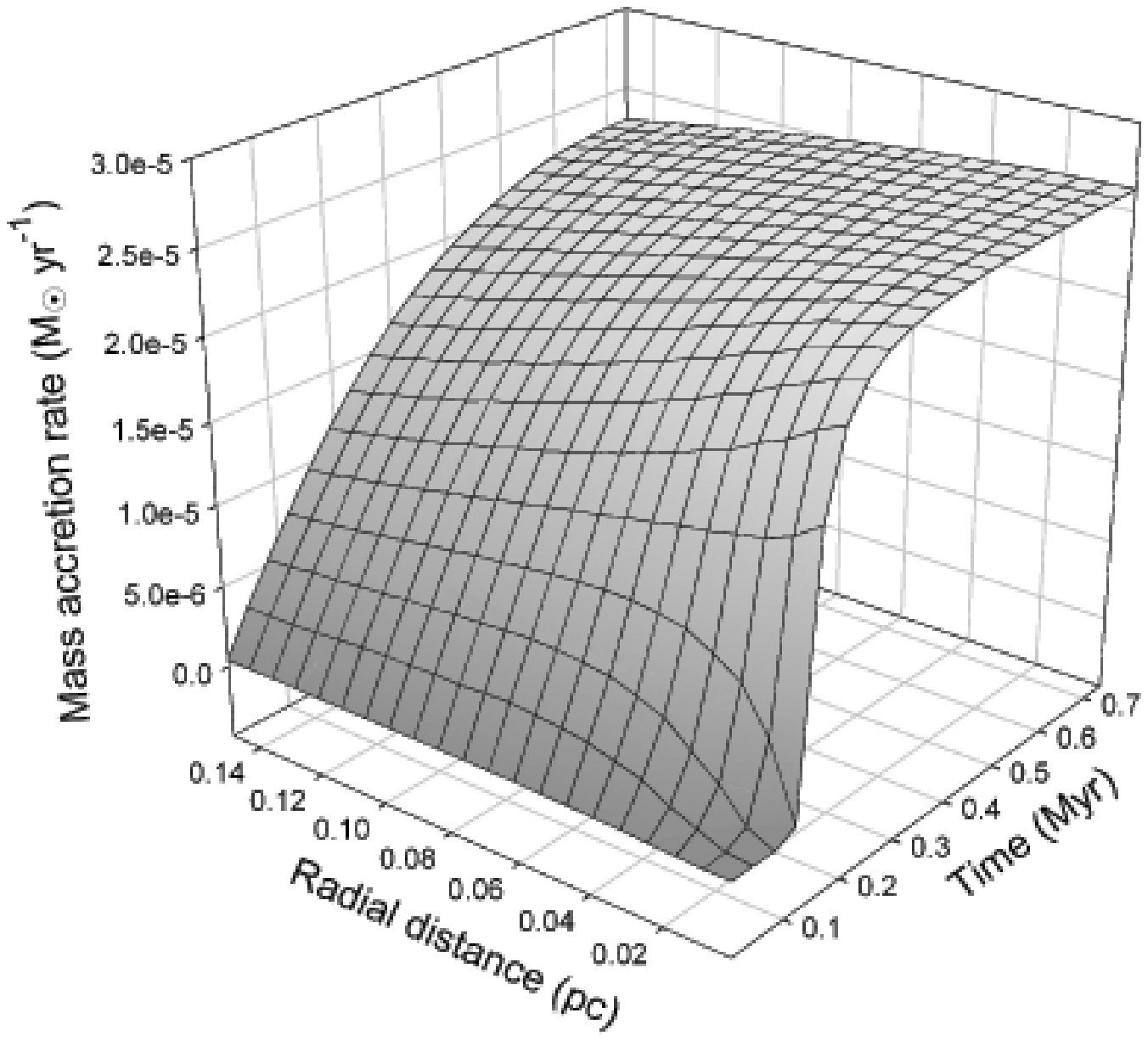}
      \caption{The mass accretion rate as a function of time and radial
      distance from the center of a pressure-free cloud that has initial radial gas density
      distribution $\rho=\rho_{\rm c}/[1+(r/r_{c})^2]$, in which $\rho_{\rm
      c}=7.5 \times 10^4$~cm$^{-3}$ and $r_{\rm c}=0.033$~pc.}
         \label{fig7}
\end{figure}

\subsubsection{Modified isothermal sphere}
First, we consider the radial gas density profile of a modified isothermal
sphere: $\rho=\rho_{\rm c}/[1+(r/r_{c})^2]$ (Binney \& Tremaine 
\cite{BT}), where
$\rho_{\rm c}$ is the gas density in the center of a cloud and
$r_{\rm c}$ is the radial scale length.
Figure~\ref{fig7} shows $\dot{M}(r,t)$ of a  pressure-free
cloud with $\rho_{\rm c}=5.5\times 10^4$~cm$^{-3}$ and $r_{\rm c}=0.033$~pc.
The mass accretion rate increases with time
and appears to approach a constant value at later times $t>0.7$~Myr.
Note that the temporal evolution of the mass accretion rate depends on
the radial distance $r$: $\dot{M}$ approaches faster a constant value
at smaller $r$.  This behavior of $\dot{M}(r,t)$
is independent of the adopted values of $\rho_{\rm c}$ and $r_{\rm c}$.
\begin{figure}
\includegraphics[width=84mm]{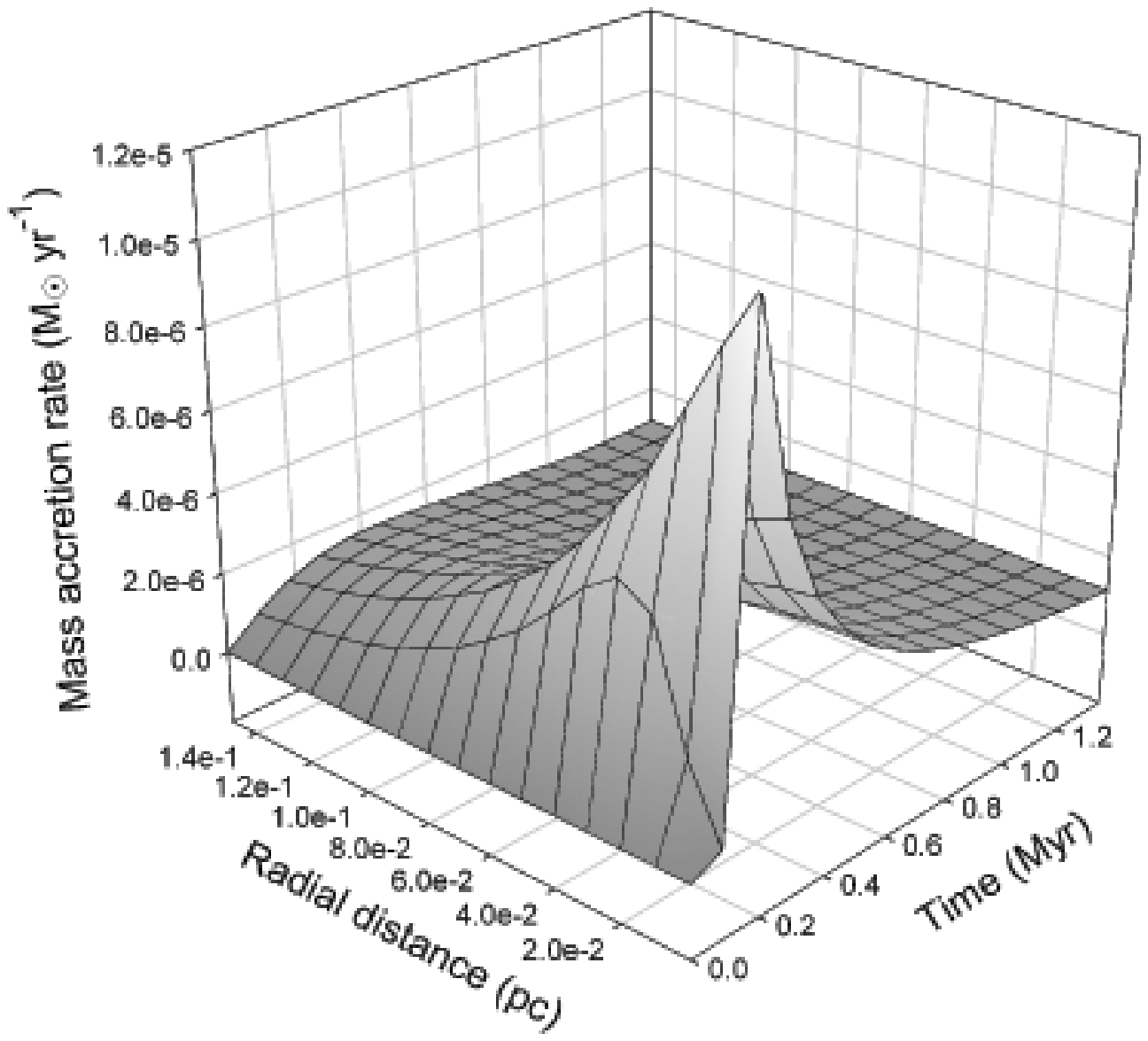}
      \caption{The same as Fig.~\ref{fig7} but for the initial gas density
      distribution $\rho=\rho_{\rm c}/[1+(r/r_{c})^3]$. }
         \label{fig8}
\end{figure}
{\sl }
\subsubsection{A steeper profile}
The submillimeter and mid-infrared observations of Ward-Thompson
et al. (\cite{WT}) and Bacmann et al. (\cite{Bacmann}) suggest
that the gas density in the envelope of a starless core falls off
steeper than $r^{-2}$. As a second example, we consider a pressure-free cloud with the
initial gas density profile $\rho=\rho_{\rm c}/[1+(r/r_{\rm c})^3]$ 
and plot the corresponding mass accretion rate $\dot{M}(r,t)$
in Fig.~\ref{fig7}. The values
of $\rho_{\rm c}$ and $r_{\rm c}$ are retained from the previous example.
As is seen,
the temporal evolution of $\dot{M}$ strongly depends on the radial distance
$r$. At $r\la10^4$~AU, the mass accretion rate has a well-defined
maximum at $t \approx 0.21$~Myr, when
the central gas density has exceeded $10^{10}$~cm$^{-3}$ (the central
stellar core
formation). After stellar core formation, $\dot{M}$ drops as $\sim t^{-1}$.
At $r>10^4$~AU, the temporal evolution of $\dot{M}$ does not show a
well-defined maximum.

\end{document}